\documentclass[aps,twocolumn,float,floatfix,prl,nofootinbib]{revtex4} 
\usepackage{graphics,graphicx,epsfig, sidecap} 
\usepackage{amssymb,xcolor} 
\usepackage{epsf,epstopdf,wrapfig} 
\usepackage{amsmath} 
 \usepackage{textgreek,bm}
\usepackage{sidecap} 
\usepackage{bbm}
\usepackage{comment}
\newcommand{\beq}{\begin{equation}} 
\newcommand{\eeq}{\end{equation}} 
\newcommand{\beqn}{\begin{eqnarray}} 
\newcommand{\eeqn}{\end{eqnarray}}

\begin{document} 

\title{Optimization and variability can coexist} 

\author{Marianne Bauer,$^{a,b,d}$ William Bialek,$^{a,b}$ Chase Goddard,$^a$ Caroline M.~Holmes,$^a$\\ Kamesh Krishnamurthy,$^{a,c}$ Stephanie E.~Palmer,$^e$ Rich Pang,$^{a,c}$ David J.~Schwab,$^f$ and Lee Susman$^{a,g}$}

\affiliation{$^a$Joseph Henry Laboratories of Physics, $^b$Lewis--Sigler Institute for Integrative Genomics,  and $^c$Princeton Neuroscience Institute, Princeton University\\
%, Princeton, NJ 08544 USA\\
$^d$Department of Bionanoscience, Kavli Institute of Nanoscience, Technische Universiteit Delft\\
%, Van der Maasweg 9, 2629 HZ Delft, the Netherlands\\
$^e$Department of Organismal Biology and Anatomy and Department of Physics, The University of Chicago\\
%, Chicago IL 60637 USA\\
$^f$Initiative for the Theoretical Sciences, The Graduate Center, City University of New York\\
%, 365 Fifth Ave, New York NY 10016 USA\\
$^g$Sagol School of Neuroscience, Tel Aviv University}
%, Tel Aviv 69978, Israel} 

\date{\today}

\begin{abstract}
Many biological systems perform close to their physical limits, but promoting this optimality to a general principle seems to require implausibly fine tuning of parameters. Using examples from a wide range of systems, we show that this intuition is wrong. Near an optimum, functional performance depends on parameters in a ``sloppy'' way, with some combinations of parameters being only weakly constrained.  
Absent any other constraints, this {\em predicts} that we should observe widely varying parameters, and we make this precise:  the entropy in parameter space can be extensive even if performance on average is very close to optimal. This removes a major objection to optimization as a general principle, and rationalizes the observed variability.
\end{abstract}

\maketitle

%\section{Introduction}

Humans can count single photons on a dark night \cite{rieke+baylor_98}, insects see the world with a resolution close to the limits set by diffraction through the lenses of the compound eye \cite{barlow_52,snyder_77}, and bacteria navigate chemical gradients with a precision limited by random arrival of molecules at their surface receptors \cite{berg+purcell_77}.  In these and other examples, living systems approach fundamental physical limits to their performance, and in this sense are close to optimal.  In many different contexts it has been suggested that this sort of optimization should be elevated to a principle from which the behavior and underlying mechanisms of these complex systems can be derived \cite{bialek_12,bialek_24}.   In contrast, others have argued that evolution cannot find optimal solutions, and emphasize examples in which the assignment of complex structures and mechanisms to functional behavior is misleading \cite{gould+lewontin_79}.  One important observation is that the parameters of biological systems, from the copy numbers of proteins in single cells to the strengths of synapses between neurons, can be highly variable \cite{spudich+koshland_76,prinz+al_04,marder+goaillard_06}, and this seems to be direct evidence against the idea that these parameters have been tuned to optimal values.

\begin{table*}
\begin{center}
\begin{tabular}{|c|c|c|}
\hline
{\bf system} & {\bf performance} ${\cal F}$ & {\bf parameters}  $\bm{\theta}$\\\hline\hline
photoreceptor array & information about visual input & receptor positions  \\\hline
transcription factors in fly embryo & positional information & readout thresholds\\\hline
small neural circuit (STG) & match to desired period or duty cycle & channel protein copy numbers\\\hline
recurrent neural network & match to desired filter & synaptic strengths\\\hline
deep artificial network & image classification performance & synaptic strengths\\\hline\end{tabular}
\end{center}
\caption{Summary of the systems used to explore how optimizing a performance function, which we can think of as a surrogate for fitness, impacts variation in system parameters. Each of the five examples show variation near their optima, but in different ways across a wide range of problems and solutions.
\label{problems}}
\end{table*} 

Intuitions about the ability of evolution to find optima and about the significance of parameter variations both depend on hypotheses about the landscape for optimization.  If optima are small, sharply defined regions in a rugged terrain \cite{wright_32, szendro+al_13}, then plausible dynamics for the exploration of high dimensional parameter spaces are unlikely to find the optimum;  similarly, even small variations in parameters would drive the system far from optimality.  We will see that this picture of the landscape is wrong in many cases.  Instead, the dependence of functional performance on the underlying parameters is very gentle or ``soft,'' so that at least some combinations of parameters can vary substantially with very little effect.   As a result, many of the arguments against optimization lose their force.  In particular, in the absence of any other constraints, a theory of systems that are very close to optimal {\em predicts} that we should observe widely varying parameters. We try to make this precise, exploring several different biological systems as well as biologically inspired model neural networks.

Our analysis is inspired in part by the idea of sloppy models \cite{transtrum+al_15}.  When we fit multi--parameter models to the observed behavior of complex systems, we often find that some combinations of parameters are determined precisely, and others are not \cite{brown+sethna_03,gutenkunst+al_07}.  Sethna and colleagues have argued that this is generic \cite{waterfall+al_06}; that the smoothness of models implies that the sensitivity to parameters is distributed more or less uniformly on a logarithmic scale, so that the fit is orders of magnitude more sensitive to some combinations of parameters than to others \cite{quinn+al_19};   and that this behavior provides a path to model simplification \cite{quinn+al_22}.   

To address the issue of optimization in living systems, we consider not the quality of a model's fit to data, but the functional performance of the organism at some task essential for its survival. The parameters are not something we choose, but rather the physical properties of the organism's internal mechanisms---concentrations of molecules, binding constants, positions of cells, strengths of synapses---that are adjusted through adaptation, learning, and evolution.   Sloppiness or the existence of soft modes then means that organisms can be very close to optimal performance while its physical properties can vary along many dimensions.    

It is tempting to think that soft modes appear just because there are many parameters feeding in to a single functional output, so that there is a kind of redundancy.  While this plays a role, we will see that soft modes appear in addition to any true redundancies. We also will see that these modes emerge in part from the structure of the problem that needs to be solved by the biological system rather than being generic features of the underlying mechanisms.

In analyzing complex models, sloppiness suggests a path to simplification. Similarly, it might seem that mechanisms with parameters that have little effect on functional performance are somehow inefficient, but this notion of parameter efficiency then is something we should have added to the performance measure from the start.  We also will see that soft modes in parameter space generally are not aligned with the microscopic components; for example, sloppiness of a genetic network crucial for fly development does not mean that we can knock out genes and maintain the same level of functional performance, even after readjustment of all other parameters \cite{sokolowski+al_24}.  These and other considerations mean that soft modes or sloppiness have consequences for the organism, not just for our models.

Evolution, learning, and adaptation allow organisms to explore their parameter spaces.  A natural hypothesis is that this exploration ranges as widely as possible while maintaining, on average, some high level of performance.  Our results on the spectrum of soft modes allows us to turn this hypothesis into quantitative predictions about observable variations in parameters, especially in the limit that parameter spaces are very high--dimensional.  Quantitatively, we argue that the entropy of the allowed variations in parameter space can be extensive even if performance is on average arbitrarily close to optimal.  Thus variability is not a retreat from optimization to a weaker notion of being ``good enough;'' rather it is a feature of these systems that optimization and variability coexist.

\section{Formulating the problem}

In different biological systems, the relevant measure of functional performance is different. If we can think about the organism as a whole in the context of evolution, then by definition the relevant measure is fitness.  In many cases we want to think more explicitly about particular systems within the organism to which we can ascribe less abstract goals.  In a neural circuit that generates a rhythm what might be important is the frequency of the rhythm; in a developing embryo what might matter is the ability of cells to adopt fates that are appropriate to their positions;  for an array of receptor cells the measure of performance might be the amount of sensory information that these cells can capture from the outside world.
In the same spirit, the underlying parameters of different systems are different.  The dynamics of a neural circuit are determined by the number of each different kind of ion channel in each cell and by the strengths of the synapses between cells;  the ability of a genetic network to read out positional information in the embryo is determined in part by the thresholds at which different genes are activated or repressed;  the information gathered by receptor arrays depends on the positions of the cells in the sensory epithelium.
We want to address these different problems in a common language.

We will write the measure of functional  performance as $\cal F$, where larger $\cal F$ means better performance.  We write the parameters of the underlying mechanism in a vector $\bm{\theta} \equiv \{\theta_1 ,\, \theta_2,\, \cdots ,\, \theta_K\}$, where the number of parameters $K$ may be quite large.  There is some function ${\cal F}(\bm{\theta})$ that determines how performance depends on parameters, and there is some parameter setting $\bm{\theta}  =\bm{\theta}^*$ that maximizes performance.  In the neighborhood of the optimum we can write
\begin{equation}
{\cal F}(\bm{\theta}) = {\cal F}_{\rm max} - \frac{1}{2}   \sum_{{\rm i}, {\rm j}  =1}^K H_{\rm ij} (\theta_{\rm i} - \theta_{\rm i}^*) (\theta_{\rm j} - \theta_{\rm j}^*)  +\cdots ,
\label{harmonic}
\end{equation}
where the Hessian matrix
\begin{equation}
H_{\rm ij} \equiv - \frac{\partial^2 {\cal F}(\bm{\theta})}{\partial\theta_{\rm i}\partial\theta_{\rm j}}{\bigg |}_{\bm{\theta}  =\bm{\theta}^*} .
\label{Hessian1}
\end{equation}
The eigenvectors of the Hessian define combinations of parameters that have independent effects on the performance, and the corresponding eigenvalues $\lambda_\mu$ measure the strength of these effects. If we think of the (negative) performance as an energy function then the eigenvalues measure the stiffness of springs that hold the parameters to their optimal setting; small eigenvalues correspond to soft springs.
Our goal is to understand the spectrum of spring constants that arise in different systems, using the examples summarized in Table \ref{problems}.

\section{Sampling the visual world}

\begin{figure}[t]
\includegraphics[width = \linewidth]{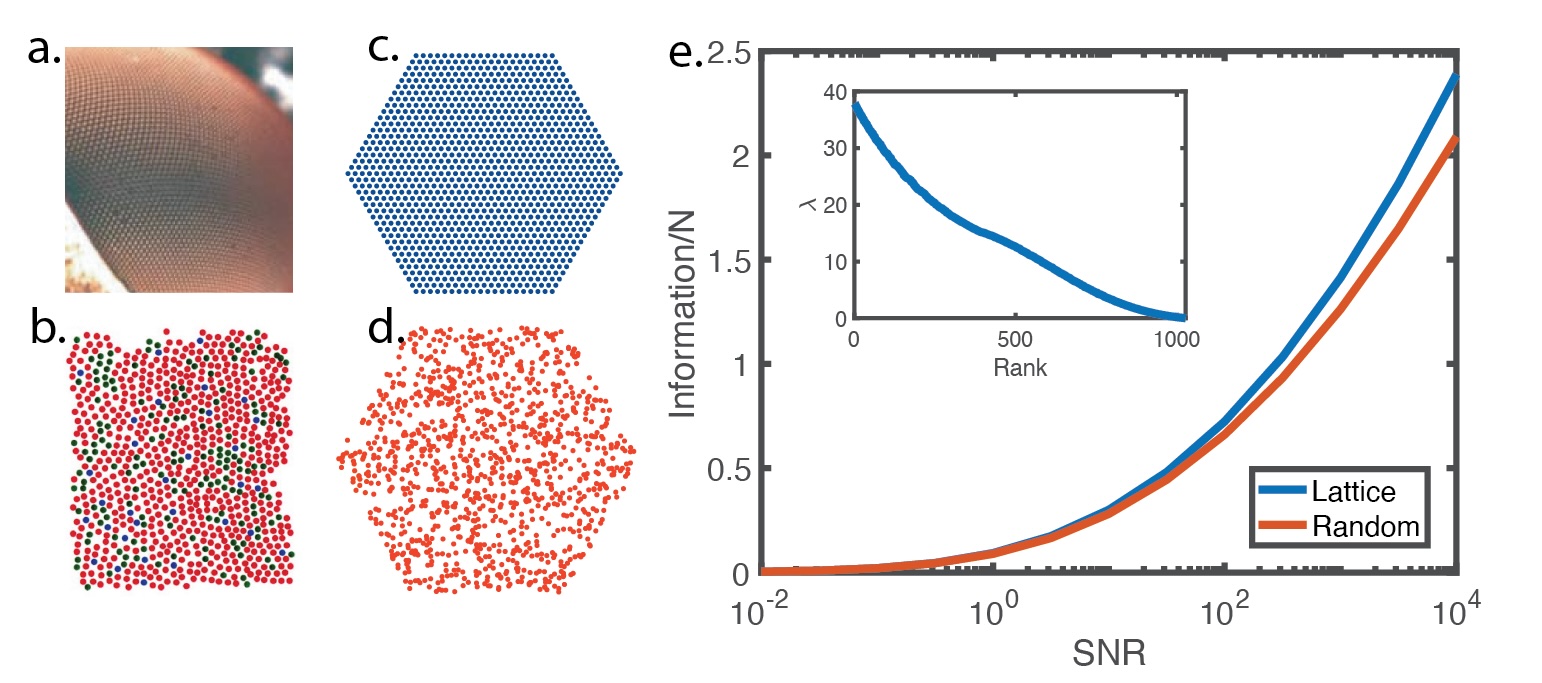}
\caption{Sampling the visual world. (a) A segment of the fly's eye showing the regular lattice of lenses, which is echoed by a regular lattice of receptors and processing circuitry \cite{Leertouwer_1991}. (b) Photoreceptor array in a human retina, with cells colored by their spectral sensitivity; note the more random arrangement \cite{roorda+al_01}. (c) Array of $N=1141$ receptors on a regular lattice.  (d) The same number of receptors scattered at random.  (e) Functional performance---here, information per cell---from Eqs (\ref{Ireceptor}, \ref{Cnm}), as a function of the signal--to--noise ratio for individual cells in the regular (blue) or random (red) arrangements. The range of  $SNR$ is chosen to match estimates for human cones in moderate daylight \cite{ruderman+bialek_94}. Inset shows the spectrum of eigenvalues of the Hessian of the functional performance, from Eq~(\ref{Hessian1}).
\label{receptors}}
\end{figure}

The lenses of the insect compound eye have a nearly crystalline arrangement, as seen in Fig~\ref{receptors}A, and this structure is repeated through several layers of visual processing circuitry \cite{strausfeld_76}.  It is an old idea that this structure optimizes the image forming or information gathering capacity of the eye \cite{barlow_52,snyder_77}.  More generally, we expect that regular sampling will maximize the capture of information, which suggests that the retinae of humans and other vertebrates---with rather random packing of photoreceptor cells, as in Fig~\ref{receptors}B \cite{roorda+al_01, szel1996distribution, sherry1998identification, jiao2014avian}
---must be far from optimal.  Perhaps surprisingly, this is not true.

In this problem, we will take the functional measure of performance to be the information that the array of $N$ receptor responses convey about the visual input, normalized by the number of cells, so that ${\cal F} = I/N$. The parameters  are the positions of the cells on the retina, $\bm{\theta} =\{\vec x_1,\, \vec x_2,\, \cdots ,\, \vec x_N\}$, so the dimensionality of the parameter space $K=2N$.  To calculate ${\cal F}(\bm{\theta})$ we need to describe the statistics of the visual input and the nature of cells' responses.  As explained in Box 1, we assume that receptors respond linearly to the light intensity as seen through an optical blurring function, and that input images are described by the power spectrum of spatial fluctuations in contrast measured in natural scenes \cite{ruderman+bialek_94}.  

In Figure \ref{receptors} we see an array of $N = 1141$ receptors arranged in a regular lattice (Fig~\ref{receptors}C)  or completely at random (Fig~\ref{receptors}D) across the same area.  The resulting information per cell is shown as a function of $SNR$ in Fig~\ref{receptors}E. The lattice solution outperforms all random solutions; however, it does so only narrowly across all signal--to--noise ratios.  This implies that the space of near-optimal retinal arrays is enormous and is on a scale similar to the entire volume of the configuration space. 

\medskip
\noindent\fbox{\begin{minipage}[c]{\dimexpr\linewidth-2\fboxsep-2\fboxrule\relax}
{\bf Box 1: Information in a receptor array}\\
\\
The retina samples the image intensity or contrast $\phi ({\vec x})$, which varies across position ${\vec x}$,  with $N$ receptors centered at locations ${\vec x}_{\rm n}$, ${\rm n} = 1,\, 2,\, \cdots ,\, N$.  This sampling involves local spatial averaging, at a minimum from the optics of the eye's lens(es), and the receptor response $R_{\rm n}$ will be noisy, so we can write
\begin{equation}
R_{\rm n} = \int d^2 x \, f({\vec x} - {\vec x}_{\rm n}) \phi({\vec x}) + \xi_{\rm n}.
\label{receptor_model}
\end{equation}
For simplicity we will assume that both signal and noise are Gaussian, with the noise independent in each receptor, $\langle \xi_{\rm n}\xi_{\rm m}\rangle = \delta_{\rm nm}\sigma^2$,
and the image statistics described by their power spectrum $S(\vec k )$.  Measurements on natural scenes show that the spectrum is scale invariant \cite{ruderman+bialek_94}, with
\begin{equation}
S(\vec k ) = \frac{A}{|\vec k |^{2-\eta}} ,
\end{equation}
and  $\eta \sim 0.2$. We treat the spatial averaging as Gaussian, $f({\vec x} ) = ({1}/{2\pi x_0^2}) \exp\left[ -  {|{\vec x}|^2}/{2 x_0^2} \right]$,
and assume that the width of this blur is matched to the overall density of receptors $\rho = 1/x_0^2$.
In this setting we find that the mutual information between the set of responses $\{R_{\rm n}\}$ and the image $\phi (\vec x ) $ is
\begin{equation}
I = \frac{1}{2} \log_2 \det \left(  \mathbbm{1}  + \hat C \right),
\label{Ireceptor}
\end{equation}
and the $N\times N$ matrix $\hat C$ has elements
\begin{equation}
C_{\rm nm} = \frac{A}{x_0^\eta\sigma^2}\int_0^\infty \frac{dz}{2\pi} 
\frac{e^{-z^2}}{z^{1-\eta}} J_0(zr_{\rm nm}/x_0) ,
\label{Cnm}
\end{equation}
where $J_0$ is a Bessel function and $r_{\rm nm} = |{\vec x}_{\rm n} - {\vec x}_{\rm m}|$.  We can choose $x_0$ as the unit of distance, so the only free parameter is a dimensionless signal--to--noise ratio $SNR = C_{\rm nn}$. A natural performance measure is the information per receptor cell, ${\cal F}= I/N$.
\end{minipage}}
\medskip

Figure \ref{receptors}E provides a global view of the space available for parameter variation with rather small changes in the functional performance of the system.  We can give a local analysis, computing the Hessian as in Eq (\ref{Hessian1}), with the spectrum of eigenvalues shown in the inset to Fig~\ref{receptors}E, around the optimal crystalline arrangement.   The eigenvectors of the Hessian are waves of receptor displacements across the array, as expected from translation invariance with corrections for the boundaries.  The eigenvalues scale roughly as $\lambda \sim 1/\ell^2$ for long wavelengths $\ell$, giving a finite density of soft modes, close to $\lambda = 0$, for parameter variation.  This means that performance is only a weakly varying function of the parameters, consistent with the result in Fig~\ref{receptors}E that even random parameter choices are not so far from optimal.

These results allow us to understand how the human retina can provide near optimal gathering of visual information while being disordered, allowing different instantiations of disorder in different individuals.  In this precise sense, optimality and variability can coexist.  In the compound eye, the size of the lens is the distance between receptors; formally this would be described in Eq~(\ref{receptor_model}) by having a different blurring function $f({\vec x})$ for each cell, with a width related to the distances to neighboring cells.   But changing the size of the lens has a large effect on the degree of blur at each point in the retina \cite{barlow_52,snyder_77},  so that disorder in the receptor lattice would drive each lens away from its optimal size.   Because of these local constraints on the optics, optimality and variability cannot coexist in compound eyes the way they do in vertebrate eyes, and correspondingly we see near crystalline structures.

\section{Information and transcriptional regulation}

Levels of gene expression provide information about variables relevant in the life of the organism.  In a developing embryo, for example, it is an old idea that concentrations of crucial ``morphogen'' molecules carry information about position, driving cell fate decisions that are appropriate to their locations \cite{wolpert_69}, and this can be made precise \cite{dubuis+al_13,tkacik+al_15,tkacik+gregor_21}. In the fruit fly all of the relevant molecules have been identified \cite{lawrence_92,ew+cnv_16}, and the concentrations of just four ``gap gene'' products are sufficient to determine position with an accuracy of $\sim 1\%$, comparable to the precision of downstream events \cite{dubuis+al_13,petkova+al_19}.

There is a classical view of gap gene expression as occurring in domains that are essentially on or off \cite{jaeger2011gap}.  If we interpret the expression levels in this way, as binary variables, then much of the information about position is lost, which means that cells must ``measure'' intermediate levels \cite{dubuis+al_13}.  But any realistic mechanism for cells to respond to gap gene expression will be noisy.   This points to an optimization problem:  Given a limited capacity to measure the concentrations of the relevant molecules, how can cells use this capacity to best capture information about position \cite{bauer+al_21,bauer+bialek_23}?  

\begin{figure}[b]
\includegraphics[width = \linewidth]{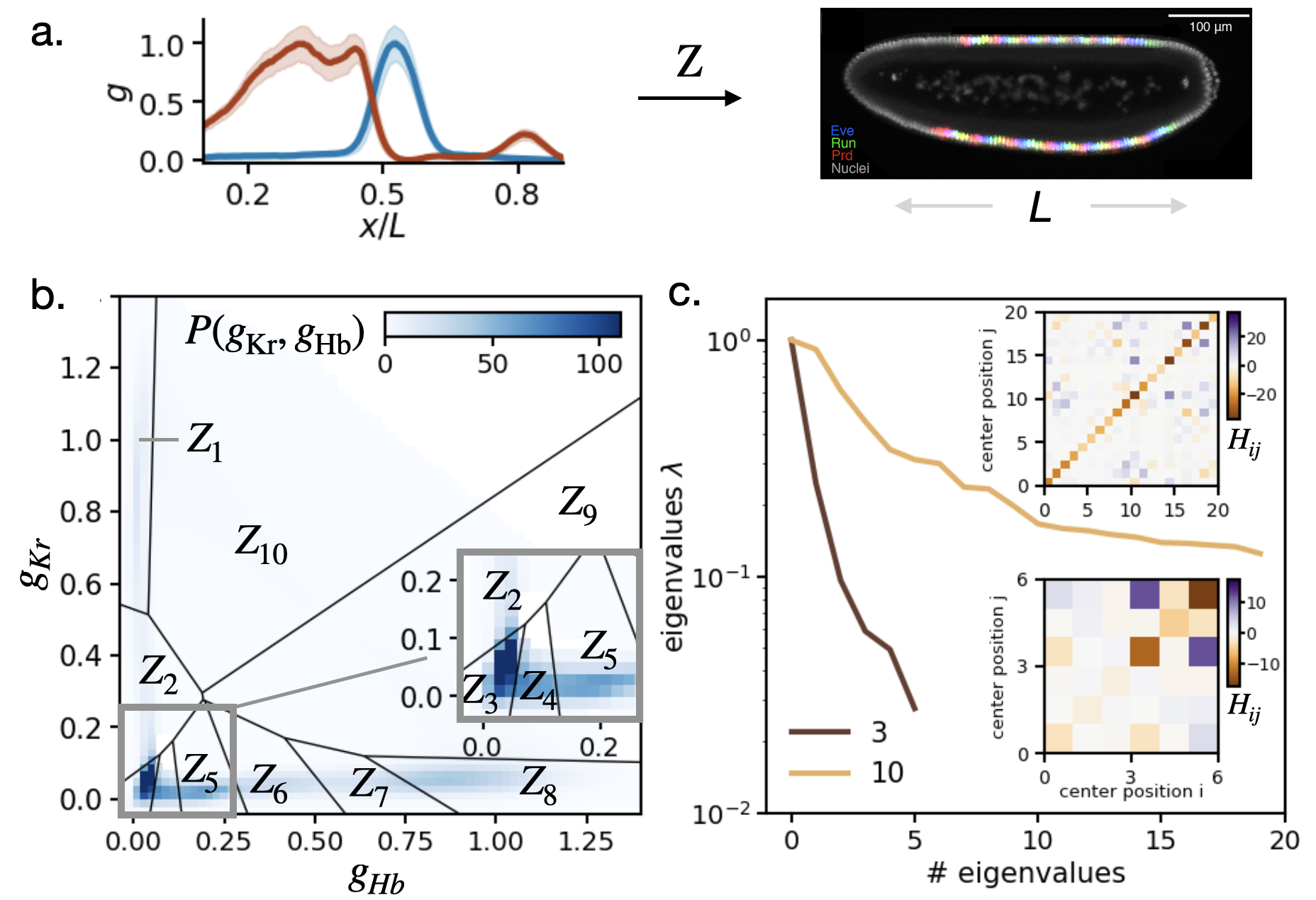}
\caption{Optimizing the response to morphogens.  (a) (left) Expression levels of Hb (red) and Kr (blue) as a function of position along the anterior--posterior axis of the embryo \cite{petkova+al_19}. Solid lines are means over many embryos, and shaded regions show the standard deviation at each position.  Expression of each gene is normalized so that $0 \leq \langle g \rangle \leq 1$, and position $x$ is measured in units of the embryo length $L$. 
(right) An image of three ``pair--rule''  genes \cite{mcgough+al_2024}, whose expression depends on enhancers reading information from all gap genes and hence on the (hypothetical) compressed variable $Z$. (b) Discretizing combinations of expression levels.  This is the optimal solution for $||Z||=10$, with regions bounded by solid lines.  Grey levels illustrate the probability density across all embryos and all positions.  (c) Hessian matrices (insets), from Eq~(\ref{Hessian1}),  and their eigenvalues for  $||Z|| =3$ and $||Z|| =10$. The peak is normalized to $1$.  
\label{fig:enhancer}}
\end{figure}

We can think of a limited capacity as being equivalent to discretization, mapping continuous gene expression values $\bm{g}$ into a small set of alternatives $Z = 1,\, 2,\, \cdots ,\, ||Z||$.  The functional performance measure is the information that this discrete variable provides about a cell's position $x$ in the embryo, ${\cal F} = I(Z; x)$, and the parameters $\bm{\theta}$ are the locations of the thresholds or boundaries that define these discrete alternatives.  Importantly, if we have enough samples from the joint distribution of $\bm{g}$ and $x$ then we can compute ${\cal F}(\bm{\theta})$ directly from data.  In Figure \ref{fig:enhancer} we do this for the genes {\em hunchback} and {\em Kr\"uppel} in the early fruit fly embryo; for this pair of genes we know that this simple discretization is a very good approximation to a more general problem of preserving positional information through limited capacity measurements of the gene expression levels (Box 2).

Figure \ref{fig:enhancer}A shows the spatial profiles 
of {\em hunchback} and {\em Kr\"uppel} in a small time window $40-45\,{\rm min}$ into the fourteenth cycle of nuclear division in the embryo.  Figure \ref{fig:enhancer}B shows the optimal discretization into $||Z||=10$ regions. Figure \ref{fig:enhancer}C shows the eigenvalues of the Hessian in the neighborhood of this optimum, both for $||Z||=3 \textrm{ and } 10$.  We see that these eigenvalues are distributed over a decade or more. We emphasize that this emerges directly from the expression level data---mean spatial profiles and the (co)variances of fluctuations around these means---and thus represents a feature of the problem faced by the organism, not a model.

\onecolumngrid
\vspace{\columnsep}

\noindent\fbox{\begin{minipage}[b]{\dimexpr\linewidth-2\fboxsep-2\fboxrule\relax}
{\bf Box 2: Compression of expression levels}\\
\\
Cells in the fly embryo respond to the proteins encoded by {\em hunchback} and {\em Kr\"uppel} as they bind to enhancer elements along the genome, and these in turn interact with promoters to influence the expression of other genes \cite{furlonglevine_18}.  We can think of this abstractly as a mapping from concentrations $g_{\rm Hb}, g_{\rm Kr}$ into some internal variable $Z$, which might represent the occupancy of enhancer binding sites or the resulting activity of the promoter \cite{bauer+al_21,bauer+al_21arxiv}.  Noise and limited dynamic range in the cellular response imply that this variable captures only limited information $I(Z; g_{\rm Hb}, g_{\rm Kr})$.  For each value of this information there is a maximum information that can be represented about the position $x$ of the cell, $I(Z;x)$, and this is determined by the structure of expression profiles and their variability.  Formally this problem can be phrased as
\begin{equation}
\max_{P(Z|g_{\rm Hb},g_{\rm Kr})} \left[ I(Z;x) - \mu I(Z; g_{\rm Hb}, g_{\rm Kr})\right],
\end{equation}
where $\mu$ is a Lagrange multiplier and in general the mapping $(g_{\rm Hb}, g_{\rm Kr}) \rightarrow Z$ can be probabilistic;  this is an instance of the information bottleneck problem \cite{tishby+al_99}.  We have solved this optimization problem \cite{bauer+al_21}, and a very good approximation is to divide the space of expression levels into  Vornoi polygons, deterministically (Fig \ref{fig:enhancer}B).  The Vornoi construction is  parameterized by the positions of the central points
\begin{equation}
\bm{\theta} = \{ \bm{g}_1 ,\, \bm{g}_2,\, \cdots ,\, \bm{g}_{N}\} ;
\end{equation}
each $\bm{g}_n$ is a vector with two components, one for {\em hunchback} and one for {\em Kr\"uppel}, so that  $\bm{\theta}$ has $||Z|| = 2N$ dimensions. Capturing more information defines a more reliable and complex body plan \cite{bialek_24}, so we take ${\cal F}(\bm{\theta})  = I(Z;x)$ as a measure of functional performance.
\end{minipage}}
\vspace{\columnsep}
\twocolumngrid

\section{Electrical dynamics of single neurons}

\begin{figure}[b]
\includegraphics[width = \linewidth]{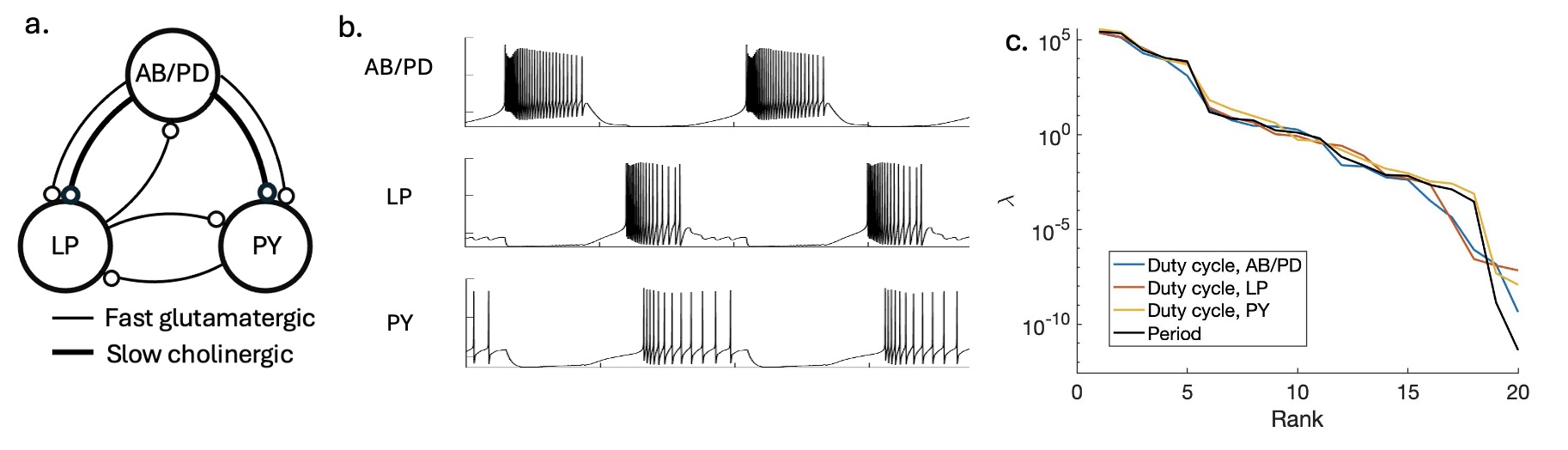}
\caption{Ion channel copy numbers in a small neural circuit. (a) Schematic of the core rhythm generating network in the crab stomatogastric ganglion \cite{prinz+al_04}, showing the anterior burster and pyloric dilator neurons (AB/PD), the lateral pyloric neuron (LP) and the pyloric neurons (PY), as well as their synaptic connections. (b) Voltage vs time for the three neurons, from the model described in Box 3. (c) Eigenvalues $\lambda$ of the Hessian in Eq (\ref{Hessian1}), where the functional performance is measured by the fractional closeness of the rhythm period (black) or duty cycle of the individual neurons (blue, green, yellow) and the parameters are the copy numbers of the different channel types in each cell.
\label{fig:channels}}
\end{figure}

The electrical dynamics of neurons are determined by the molecular dynamics of ion channels in the cell membrane \cite{johnston1994foundations, bevan1999mechanisms}. 
A typical neuron expresses several kinds of channels, chosen from roughly one hundred  possibilities encoded in the genome.   Channels interact with one another through the voltage across the membrane, and these are the protein networks for which we have the most precise mathematical descriptions.  The models of these networks predict that stabilizing even the qualitative behavior of neurons requires feedback from electrical activity to channel copy number \cite{lemasson+al_93},  and these feedback mechanisms now are well established.  Here we consider a small network of neurons that generate a rhythm in the crab stomatogastric ganglion, as schematized in Figs \ref{fig:channels}A and B \cite{prinz+al_04};  these coordinated oscillations create crucial feeding patterns in the crab. A description of the channel dynamics in the simplest three cell network requires twenty parameters $\bm{\theta}$ that define the number of each type of channel, and we measure functional performance $\cal F$  by the closeness of the period and duty cycle to  target values  (Box 3).

Figure \ref{fig:channels}C shows the eigenvalues of the Hessian, from Eq (\ref{Hessian1}).  Whether we focus on the period of the oscillation or the duty cycles of individual neurons, the results  are strikingly similar, and show that  eigenvalues are distributed almost uniformly  on a logarithmic scale.   The dynamic range is $\sim 10^{10}$, which means that some combinations of parameters can vary by {\em one hundred thousand times} as much as others while having the same effect on the performance.  Said another way, to hold all parameters fixed would require that variations of $\sim 10^{-10}$ are significant.

\onecolumngrid
\vspace{\columnsep}

\noindent\fbox{\begin{minipage}[b]{\dimexpr\linewidth-2\fboxsep-2\fboxrule\relax}
{\bf Box 3: Ion channel dynamics}\\
\\
Electrical dynamics in single neurons are described by generalized  Hodgkin--Huxley equations \cite{aidley_98}:
\begin{eqnarray}
    C_n\frac{dV_n}{dt} &=& - g_{n}^{\rm leak}V_n - \sum_{i}g_{ni}^{\rm max}m_{ni}^{\alpha_i} h_{ni}^{\beta_i}(V_n - {\bar V}_{ni}) \, 
    + I_n^{\rm syn} ,\label{HH1}\\
    \frac{dm_{ni}}{dt} &=& -\frac{1}{\tau_i^m(V_n)}\left[m_{ni} - {\bar m}_i(V_n)\right] ,\label{HH2}\\
    \frac{dh_{ni}}{dt} &=& -\frac{1}{\tau_i^h (V_n)}\left[h_{ni} - {\bar h}_i(V_n)\right].\label{HH3}
\end{eqnarray}
The variable $V_n$ is the voltage across the membrane of cell $n$ and $C_n$ is its capacitance.  There are many types of ion channel proteins that can insert into the cell membranes, and these are labeled by $i$; 
$m_{ni}$ and $h_{ni}$ are the ``activation'' and ``inactivation'' variables, respectively, for these channels, reflecting the states of the proteins.  Given the ionic selectivity of each channel type and the chemical potential differences of the ions across each cell membrane, current through channels of type $i$ reverses at a voltage ${\bar V}_{ni}$ in cell $n$.  If these channels are fully open the maximal conductance is $g_{ni}^{\rm max}$, proportional to the number of copies of that protein in the membrane.  When the voltage across the membrane is $V$, the equilibrium values of the activation and inactivation variables are ${\bar m}_i(V)$ and ${\bar h}_i(V)$, respectively, and relaxation toward these equilibrium values occurs with time constants $\tau_i^m(V)$ and $\tau_i^h (V)$; these voltage dependencies are well characterized for each channel type \cite{johnston1994foundations}.  Each cell has a ``leak'' conductance $g_{n}^{\rm leak}$ that pulls the voltage back toward zero.  Current is injected into each cell by synapses from other neurons,
\begin{eqnarray}
    I_n^{\rm syn} &=& \sum_k g_{nk}^s s_{k}(V_n -E^s_{nk})
    \label{Isyn} \\
    \frac{ds_{k}}{dt} &=& -\frac{1}{\tau^s_k (V_k)}\left[s_k - \bar{s}_k(V_k)\right].
    \label{Isyn2}
\end{eqnarray}  
The input current to the postsynaptic neuron $n$ depends on the momentary activation of the synapse from cell $k$, $s_k$, the maximal synapse strength $g_{nk}^s$, and the difference between the membrane potential of the postsynaptic neuron and the reversal potential of the synapse, $E_{nk}$.  The activation of the synapse relaxes on a timescale $\tau^s_k (V_k)$ toward a steady state  $\bar{s}_k(V_k )$. The functions  $\bar{s}_k(V_k )$ and $\tau^s_k (V_k)$ are modeled as in Ref \cite{abbott1998modeling}, with $\bar{s}_k(V_k )$ taking the form of a sigmoid
In small circuits as in Fig \ref{fig:channels} we find solutions to Eqs (\ref{HH1}--\ref{Isyn2}) that are periodic, so that $V_n(t) = V_n(t+T)$.  One measure of fitness is the (fractional) approach of this period to some desired value, ${\cal F} = {\cal F}_{\rm max} - (T-T_*)^2/T_*^2$.  
During the period $T$ each neuron spends part of the cycle in firing state and part of the cycle not firing. We can define a duty cycle $D$ for each neuron by first inferring the timing $t_i$ of each of the firing events within a single burst of firing, and then finding the length $\tau = t_f-t_1$ of the firing burst. Finally, we can define the duty cycle as $D = \tau/T$. Then, we define a fitness as the fractional approach of this duty cycle to a desired value for each cell $n$, ${\cal F}_n = {\cal F}_{\rm max} - (D_n-D_{n,*})^2/D_{n,*}^2$.  We take the parameters to be the channel protein copy numbers or maximal conductances,  $\bm{\theta} = \{g_{ni}^{\rm max}\}$.
Hessians were estimated through numerical simulation using the Xolotl package \cite{gorur2018xolotl}, using the ``example pyloric network'' to define  $D_{n,*}$ and $T_*$.
\end{minipage}}
\vspace{\columnsep}
\twocolumngrid

\section{A recurrent network}

\begin{figure}[t]
\centerline{\includegraphics[width = \linewidth]{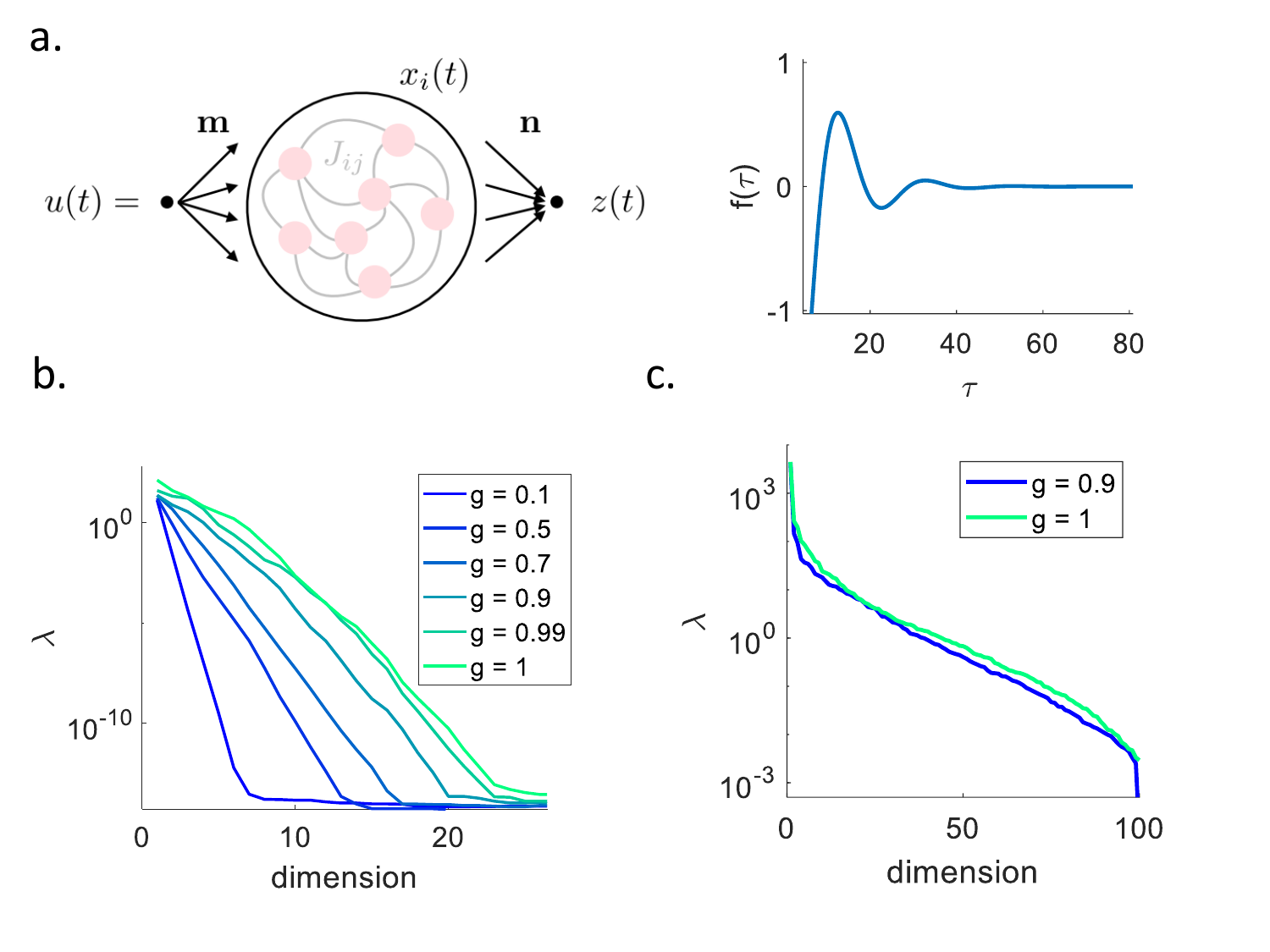}}
    \caption{A linear recurrent network of $N=100$ neurons.   (a)  Schematic  network architecture (left). A  network with internal connections $J_{\rm ij}$, is driven by a signal $u(t)$ via an input weight vector $\mathbf{m}$, and the output $z(t)$ is a weighted sum of network activity via the  readout weights $\mathbf{n}$.
    Visualization of the network filter (right), from Eq~(\ref{filt1}).
    (b) Hessian eigenvalues $\lambda$ of the cost function, Eq (\ref{eq:RNN_cost}), for a random filter $f_* (\tau )$ when perturbing the readout weights $n_i$; different values of $g$ in colors.
    (c) As in (B) when perturbing the recurrent connectivity weights $J_{\rm ij}$.     
\label{RNN}} 
\end{figure}

Going beyond single neurons to explore the dynamics of networks we can see more examples of these ideas.  We consider first a network with recurrent internal connections and a linear readout  (Fig \ref{RNN}A). 
Nonlinear versions of these networks have been widely discussed as generators  of complex outputs $z(t)$, as in motor control \cite{hennequin2014optimal, sussillo2015neural, susman2021quality}.  A much simpler case is where the dynamics are linear  and there is a single input $u(t)$ (Box 4), in which case the output  is a linearly filtered version of this signal,
\begin{equation}
z(t) = \int dt' f(t-t') u(t').
\label{filt1}
\end{equation}
We can take as a measure of functional behavior $\cal F$ the closeness of this filter to some desired target $f_*(\tau)$.  The parameters are the strengths of internal synapses and the readout weight,  $\bm{\theta} = \{ J_{\rm ij}; n_{\rm i}\}$.

We can make a generic choice of the target filter $f_*(\tau )$ by setting all the parameters to random values. Note that as we increase the variance of the synaptic strengths ($g/N$ from Box 4), the network has access to a wider range of time scales and these will be reflected in $f_*(\tau )$.  We can gain some insight by first varying only  the $N$ readout weights $\{n_{\rm i}\}$.  We find the eigenvalues  of the Hessian matrix shown in Fig \ref{RNN}B, illustrating once again  the tendency for these values to be equally spaced on a logarithmic axis.   At larger $g$, and hence a wider range of time scales,  we find a larger number of eigenvalues that are significantly nonzero, but at all values of $g$ there is a range of more than ten orders of magnitude.  Indeed, as $g\rightarrow 1$ the spectrum becomes almost perfectly uniform across this range.

When we change the $N^2$ synaptic strengths we find the eigenvalues of the Hessian shown in Fig \ref{RNN}C. To understand this result we can think not about the individual elements of the synaptic matrix $J_{\rm ij}$ but about the eigenvalues and eigenvectors of this matrix.  If perturbations to the network parameters change only the eigenvectors, then we can compensate by adjusting the readout weights.  But this means that there are $N(N-1)$ parameters of the synaptic matrix that are effectively redundant, so we see only  $N$ significantly nonzero eigenvalues.  Beyond true redundancy or over--parameterization, even the nonzero eigenvalues are widely distributed (Fig \ref{RNN}C).  These  eigenvalues span roughly five orders of magnitude, and except for a few they are distributed uniformly on a logarithmic axis, as in our other examples.

\noindent\fbox{\begin{minipage}[b]{\dimexpr\linewidth-2\fboxsep-2\fboxrule\relax}
{\bf Box 4: A linear network}\\
\\
We consider a linear network of $N$ neurons in which the activity of each neuron is described by a variable $x_{\rm i}$.  This activity relaxes to its resting state on time scale $\tau_0$, and $J_{\rm ij}$ measures the strength of the connection from neuron $\rm j$ to neuron $\rm i$. There is a single input $u(t)$ that drives each neuron $\rm i$ with a fixed weight $m_{\rm i}$, so the full dynamics are
\begin{equation}
\tau_0 \frac{dx_{\rm i}}{dt} = - x_{\rm i} + \sum_{{\rm j}=1}^N J_{\rm ij} x_{\rm j} + m_{\rm i} u(t) . \label{eq:driven_rnn}
\end{equation}
The output of the network is some linear combination of activities, $z(t) = \sum_{\rm i} n_{\rm i} x_{\rm i}(t)$.  Linearity of the dynamics implies that $z(t)$ is a linearly filtered version of $u(t)$, as in Eq~(\ref{filt1}).  We measure functional performance as the similarity of this filter to some target,
\begin{equation}
{\cal F}(\bm{\vec\theta}) = {\cal F}_{\rm max} - \int d\tau | f(\tau) - f_*(\tau)|^2 .
\label{eq:RNN_cost}
\end{equation}
We can make a generic choice of the target by drawing each synaptic strength $J_{\rm ij}$ independently from a Gaussian distribution with zero mean and $\langle (J_{\rm ij})^2\rangle = g^2/N$, and similarly picking weights $\{m_{\rm i}; n_{\rm i}\}$ from zero mean unit variance Gaussians.
In the large $N$ limit, the eigenvalues of $J_{\rm ij}$  are uniformly distributed on a disc centered at the origin of the complex plane, with radius $g$. 
Through Equation~(\ref{eq:driven_rnn}) this means that when $g\rightarrow 1$  the network dynamics have increasingly longer timescales, and when $g>1$ the dynamics destabilize. 
\end{minipage}}

\section{A deep network}

Modern artificial neural networks began as schematic models for networks of real neurons.  While not faithful to the details of the brain's dynamics, they capture important functional behaviors that are at the heart of the current revolution in artificial intelligence.  Although it often is emphasized these networks have very large numbers of parameters, networks of real neurons have even more.  Here we use deep networks \cite{lecun2015deep} as a controlled test case to further explore the interplay of optimization and variability.

Networks that are trained on classification tasks are models for a conditional distribution $Q_{\bm{\theta}} (C|\mathbf{x})$, where $C$ is the class, $\mathbf x$ is the $D$--dimensional input such as an image, and $\bm\theta$ are the (many) parameters.   The inputs themselves are drawn from a distribution $P(\mathbf{x})$.  The performance of the model can be measured by the average (log) probability that the model can explain the data we have seen, 
\begin{equation}
{\cal F}  \left(\bm{\theta} \right) = +\frac{1}{N} \sum_{{\rm i}=1}^N \log Q_{\bm\theta} (C_{\rm i}|\mathbf{x}_{\rm i}) ,
\end{equation}
where $C_{\rm i}$ is the class to which the input $\mathbf{x}_{\rm i}$ belongs.  As usual in neural networks the parameters $\bm{\theta}$ are the connection weights and biases for activation of individual units. Note that our distinction between functional performance and model fit becomes blurred in this example.

Because the network implements a probabilistic model, and the model is large enough to achieve perfect performance, the Hessian matrix of $\cal F$ is equal to the Fisher information matrix \cite{pmlr-v108-thomas20a} at the optimum, but the Fisher matrix is well defined and positive definite even away from the optimum. This allows us to ask, for example, how the soft modes emerge over the course of learning.

\begin{figure}
\includegraphics[width=\linewidth]{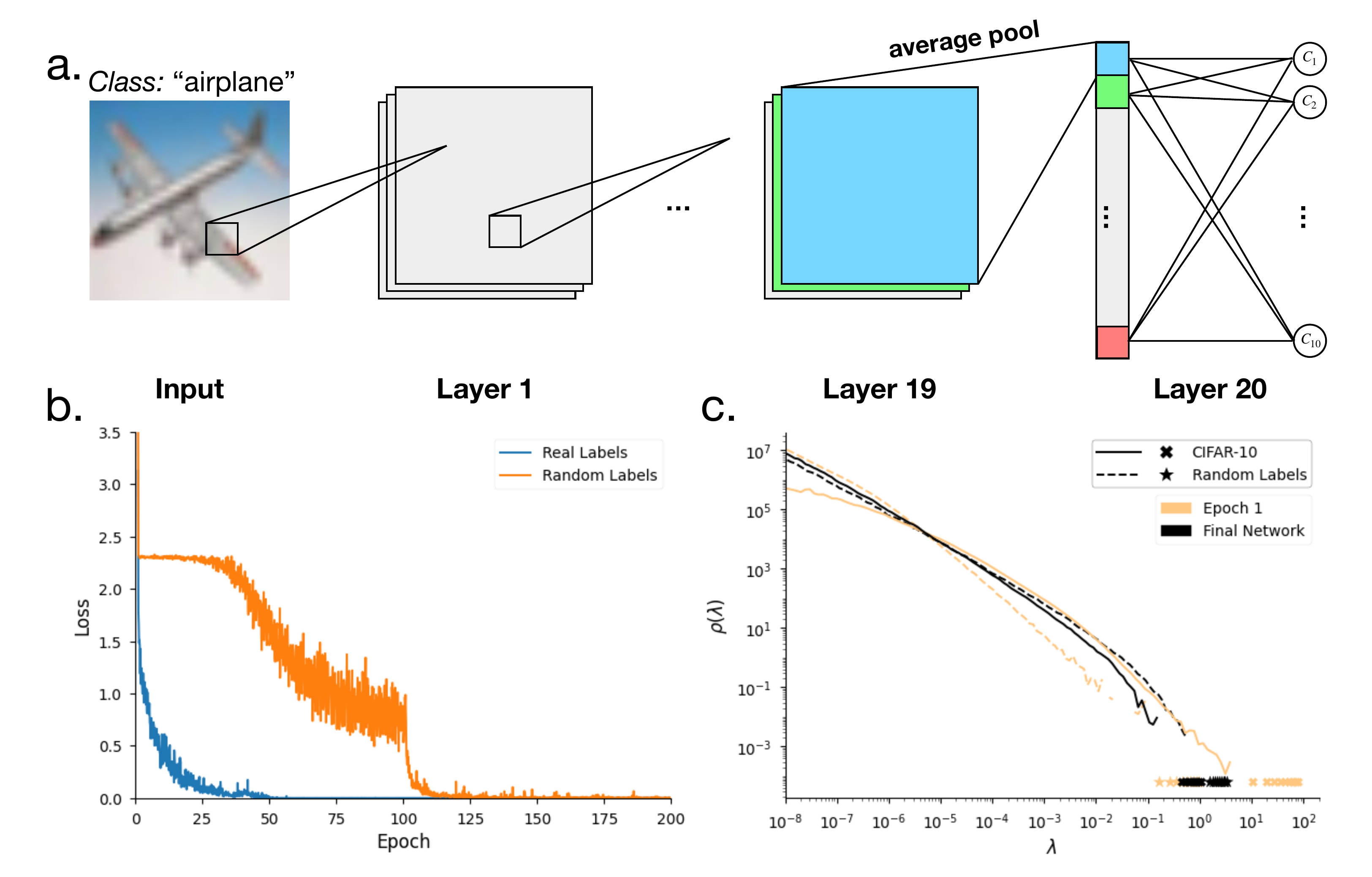}
\caption{Eigenvalue spectrum of the Fisher information matrix for a network that learns the CIFAR--10 task. (a) Schematic of the network. (b) Performance of the network as function of time (``epoch'') during training.  Training reaches a true optimum---zero loss---both for the natural task and for an artificial task in which the labels on images are randomized.   (c) Eigenvalues of the Fisher information matrix $F$ estimated from $\tilde F$ as in Eq (\ref{tildeF}).  These estimates are based on $n=1500$ samples, probing the spectrum at different stages in the learning process. The distribution of eigenvalues converges as we increase $n$, especially in the tail of small eigenvalues.\label{deep}}
\end{figure}

In most modern applications of neural networks, and perhaps in the brain itself, there are more parameters than examples from which to learn, and so we expect that some combinations of parameters are completely unconstrained, generating true zero eigenvalues in the Hessian matrix.  But there are hints in the literature that beyond these genuinely undetermined parameters there is a near continuum  soft modes \cite{sagun+al_18}.  To explore this further we consider ResNet20, a twenty layer network \cite{he+al_15,Idelbayev18a}, that is trained to $100\%$ accuracy---and hence the absolute maximum ${\cal F}$---on the CIFAR--10 task, in which $32\times32$ color images must be assigned to $|C| = 10$ classes \cite{krishevsky_09}.  A schematic of the network is shown in Fig~\ref{deep}A, and Fig~\ref{deep}B shows that learning indeed reaches the optimum, with perfect classification of the training data. This network has $K\sim 270,000$ parameters, and the Hessian matrix is $K\times K$.  As explained in Box 5, we have developed finite rank approximations to the Fisher information matrix that involve a direct average over samples. 

We see in Figure \ref{deep}C that the Hessian has just a handful of large eigenvalues; as noted  previously the number of these ``stiff'' directions is equal to the number of classes $|C|$ \cite{gur2018gradient,sagun+al_18}.  As learning progresses, the density of moderate eigenvalues is depleted and added to the tail at small eigenvalues, which becomes nearly scale invariant $\rho(\lambda ) \sim 1/\lambda$ so that the spectrum is barely integrable; there is no obvious limit to how soft these modes can be.  Importantly, the spectrum we see converges rapidly as we increase the rank of our approximation.

The classical examples of sloppy models involve eigenvalues that are distributed almost uniformly on a logarithmic scale, which is equivalent to the scale invariant distribution $\rho(\lambda) \sim 1/\lambda$ that we see here.  It is striking that we see this not only across many decades in $\lambda$, but across many thousands of dimensions.  We emphasize that this uniform distribution across scales is not a property of the generic deep network, but emerges during training, as we can see by comparing different colors in Fig~\ref{deep}C.  More subtly, if we change the task by randomly permuting the labels we can still reach optimal performance (Fig~\ref{deep}B), but in Fig~\ref{deep}C  we see that the density of moderate and large eigenvalues now grows over the course of training at the expense of the very smallest eigenvalues.

\noindent\fbox{\begin{minipage}[b]{\dimexpr\linewidth-2\fboxsep-2\fboxrule\relax}
{\bf Box 5: The Hessian in large networks}\\
\\
As before we are interested in  the  Hessian 
\begin{equation}
H_{\mu\nu} \equiv - \frac{\partial^2 {\cal F}(\bm{\theta})}{\partial \theta_\mu\partial\theta_\nu}{\bigg |}_{{\bm\theta} = {\bm\theta}^*} .
\end{equation}
At large $N$ we can replace the average over samples with an average over the true distribution $P(C|\mathbf{x})$.  But in modern deep networks, the very large number of parameters makes it possible to achieve perfect classification on the set of training examples, so that $P(C|\mathbf{x}) = Q_{{\bm\theta}^*}(C|\mathbf{x})$. When this is true, the Hessian at the optimum is the Fisher information matrix,
\begin{eqnarray}
H_{\mu\nu} &=& F_{\mu\nu} ({\bm\theta}^*)\\
F_{\mu\nu} (\theta) &=& {\bigg\langle} 
\frac{\partial  \log Q_{\bm{\theta}} (C|\mathbf{x})}{\partial \theta_\mu} 
\frac{\partial  \log Q_{\bm{\theta}} (C|\mathbf{x})}{\partial\theta_\nu} {\bigg\rangle}\\
{\bigg\langle} \cdots {\bigg\rangle} &=&  \int d^Dx \,P(\mathbf{x}) \sum_C Q_{\bm{\theta}}  (C|\mathbf{x})
(\cdots ) .
\end{eqnarray}
We note that the Fisher information matrix  is a property of the model and not of the training data, except insofar as the training drives us to a particular set of parameters $\bm{\theta}^*$.  Because $F_{\mu\nu}$ is an expectation value we can approximate it by drawing $n$ examples out of the true distribution (not necessarily examples used for training).  Concretely we can construct
\begin{eqnarray}
F_{\mu\nu}^{(n)} &=& \frac{1}{n}\sum_{{\rm i}=1}^n  \sum_C A_{\mu,({\rm i}C)} A_{\nu,({\rm i}C)} \\
A_{\mu,({\rm i}C)} &=&  
\frac{\partial  \log Q_{\bm\theta} (C |\mathbf{x}_{\rm i})} {\partial\theta_\mu}  
\end{eqnarray}
This is a rank $n|C|$ approximation to the true matrix $F_{\mu\nu}$, and  the eigenvalues of $F_{\mu\nu}^{(n)}$ are the same as the $n$ nonzero eigenvalues of 
\begin{equation}
\tilde F_{({\rm i}C),({\rm j }C')}^{(n)} = \frac{1}{n} [\hat A^T \hat A]_{\rm ij} = \frac{1}{n}\sum_{\mu =1}^K A_{\mu,({\rm i}C)} A_{\mu,({\rm j}C')} .\label{tildeF}
\end{equation}
Handling the $n|C|\times n|C|$ matrix $\tilde F$ is a manageable task for reasonably large $n$, so long as $n|C| \ll K$.  The eigenvalues of $F^{(n)}$ approach those of $F$ as $n$ increases, and we can find these eigenvalues by diagonalizing $\tilde F^{(n)}$.  If the eigenvalue spectrum of $\tilde F^{(n)}$ stabilizes as  $n$ becomes large, then we are finding a good approximation to the spectrum of $F$ itself.

\end{minipage}}

\section{Synthesis}

Across all of the systems highlighted here, we observe a large number of soft modes in parameter space.  This describes quantitatively the flexibility of parameter choices that allow equivalent and very nearly optimal performance. As parameter spaces expand, the scaling of the eigenvalues that we observe admits yet more soft modes. Our framework allows us to see the generality of these relationships across vastly different biological substrates and cost functions. 

In thinking about the possibility of parameter variation it is useful to imagine a population of cells, networks, or organisms that choose their underlying parameters $\bm{\theta}$ at random from a distribution $P(\bm{\theta})$.  This population will  exhibit an
 average performance or fitness $\bar{\cal F}$, 
\begin{equation}
\bar{\cal F} = \int d^K\theta\, P(\bm{\theta}) {\cal F}(\bm{\theta}) .
\label{barC_fix}
\end{equation}
Environmental and evolutionary pressures will push for this average to reach some characteristic level. Unless there is some mechanistic or historical constraint beyond the fitness itself, we expect that the parameters will be as variable as possible while obeying the mean value in Eq (\ref{barC_fix}).  

The only consistent way to make ``as variable as possible'' mathematically precise is to ask for the distribution $P(\bm{\theta})$ that has the maximum entropy consistent with the constraint on $\bar{\cal F}$ \cite{shannon_48,jaynes_57}.  This maximum entropy distribution has the form
\begin{equation}
P(\bm{\theta})  = \frac{1}{Z(T)} \exp\left[ + \frac{1}{T} {\cal F}(\bm{\theta})\right],
\label{boltz}
\end{equation}
where $T$ is a parameter that must be set so that Eq (\ref{barC_fix}) is satisfied and $Z(T)$ enforces the normalization of the distribution.  We recognize this as the Boltzmann distribution with the performance measure playing the role of the (negative) energy and $T$ the temperature.

A seemingly different view of the problem is that there is some added cost in controlling the parameters, and this competes with the fitness advantage that can be gained by tuning towards an optimum.  If the cost is proportional to the information, in bits, that is used to specify the parameters,
then achieving a certain mean fitness at minimal cost means that the distribution of parameters will again be drawn from Eq (\ref{boltz}).  In this picture the effective temperature $T$ measures the relative cost of the information needed for control vs the gain in fitness.

If the evolutionary pressure is strong enough to keep the mean performance close to its maximum, then we can use the expansion in Eq (\ref{harmonic}), and in agreement with the equipartition theorem, we find 
\begin{equation}
\bar{\cal F} = {\cal F}_{\rm max} - \frac{K}{2} T,
\label{fit_equi}
\end{equation}
independent of the details of the eigenvalue spectrum. We can explicitly insert our knowledge of the eigenvalue spectrum into the expression for the entropy of the distribution, 
\begin{equation}
S = -\int d^K\theta\, P(\bm{\theta}) \ln P(\bm{\theta}) = \frac{1}{2}\sum_{\mu =1}^K \ln (2\pi e T/\lambda_\mu ) .
\end{equation}
If we have a sloppy spectrum with
\begin{equation}
\lambda_\mu = \lambda_{\rm max} e^{-\alpha\mu} .
\label{sloppy}
\end{equation}
then 
\begin{equation}
S =  \frac{K}{2}  \ln(2\pi e T/\lambda_{\rm max} ) + \frac{\alpha}{4} K(K+1).
\end{equation}
In this way, we can relate the entropy to the temperature and dimensionality, $K$, of the system. Solving for $T$ and inserting the expression into Eq (\ref{fit_equi}), we can also probe the scaling of the cost function with dimensionality. Putting things together we have
\begin{equation}
\Delta{\cal F}\equiv \bar {\cal F}_{\rm max} - \bar {\cal F} = \frac{K\lambda_{\rm max}}{4\pi e} \exp\left[ \frac{2S}{K} - \frac{\alpha}{2} (K+1)\right].
\label{result1}
\end{equation}
This implies that if the the dimensionality of the parameter space is large ($K\rightarrow\infty$) we can get arbitrarily close to optimal performance, on average, even though the variability corresponds to a finite entropy per parameter.

It is useful to think about the case where parameter spaces are so large that the discrete Hessian eigenvalues are drawn from a nearly continuous distribution with normalized density $\rho (\lambda )$, as in Fig \ref{deep}.  Then the analog of Eq (\ref{result1}) is
\begin{equation}
\Delta{\cal F} = \frac{K\lambda_{\rm max}}{4\pi e} \exp\left[ \frac{2S}{K} + \int_{\lambda_{\rm min}}^{\lambda_{\rm max}} d\lambda \,\rho(\lambda ) \ln (\lambda/\lambda_{\rm max})  \right].
\label{result2}
\end{equation}
The condition to have finite entropy per parameter at vanishing fitness cost is that the dynamic range of $\ln\lambda$ grow with the dimensionality $K$ of the parameter space.  This growth is linear if the sloppy spectrum in Eq (\ref{sloppy}) is exact, but this can happen in a wider range of systems.

Similar considerations appear in many examples beyond those discussed here.  In some of these problems the parameters are discrete, as for the mapping of amino acid sequences into protein structure and function, but the principles are the same.  The fact that proteins form families is a hint that the mapping from sequence to structure is many--to--one \cite{stroud_74,pfam,blum+al_20}, and this observation is at the root of a large literature that builds statistical physics models for the distribution of sequences consistent with a given structure \cite{lapedes+al_98,bialek+ranganathan_07,weigt+al_09,marks+al_11,russ+al_20}.   Within these models one can estimate the entropy of sequence variations at fixed structure, analogous to the entropy at fixed mean performance discussed here; again this entropy appears to be extensive in the number of amino acids \cite{barton+al_16}.  This is not a generic feature of proteins, and it has been suggested that evolution selects for structures that are the stable folded states of many sequences, in effect maximizing the entropy in parameter space \cite{li+al_96,li+al_98}.
Direct experimental explorations of sequence space also suggest that the landscape may be more easily navigable than previously expected \cite{carneiro+hartl_10,papkou+al_23}.

On a higher level of organization, many cognitive tasks can be thought of as sequential decision making, leading to a natural discrete parameterization of the space of strategies. There is a regime in which tasks are complex enough to be interesting but small enough that this space can be explored exhaustively. 
For foraging tasks, this enumeration reveals a broad swath of strategies that differ significantly in form but have negligible loss in performance, very much analogous to the results above \cite{ma2024vast}.

\section{Conclusion}

Contrary to a widely held intuition, near--optimal performance of biological systems does not require fine tuning of parameters.  We have seen this in examples across scales, from the control of gene expression to the dynamics of large neural networks.  This is not a generic feature of high--dimensional, ``over--parameterized'' systems,  but rather emerges from an interplay between the underlying mechanistic architecture and the structure of the problem that the system is selected to solve. Nor is the tolerance for parameter variation a retreat from optimality to ``good enough'' solutions; rather it is a feature of the landscape for optimization itself.   Optimization principles for the function of living systems thus predict that there should be substantial variation in the underlying parameters.  We should not be surprised by optimality, but rather only by those rare cases where microscopic parameters are controlled beyond what seems necessary for function.

\section{Acknowledgments}

Portions of this work were presented at the NITMB workshop on Biological Systems that Learn, and we thank our colleagues there for helpful discussions, especially M Holmes--Cerfon. This work was supported in part by the National Science Foundation through the Center for the Physics of Biological Function (PHY--1734030) and a Graduate Research Fellowship (to CMH); by the National Institutes of Health through the BRAIN initiative (1R01EB026943) and Grant R01NS104899; by Fellowships from the Simons Foundation and the John Simon Guggenheim Memorial Foundation (WB); by a Polymath Fellowship from the Schmidt Sciences Foundation (SEP); by a Jane Coffins Childs Memorial Fund Fellowship (CMH); and by NWO Talent/VIDI grant NWO/VI.Vidi.223.169 (MB).  WB also thanks colleagues at Rockefeller University for their hospitality during a portion of this work.

\bibliography{opt+var}

\end{document}